\documentclass[newstyle,twocolumn,proceedings]{rmaa}
\renewcommand{\P}[1]{%
\ifnum#1=1\hbox{OW~168--326E}\fi
\ifnum#1=2\hbox{OW~167--317}\fi
\ifnum#1=3\hbox{OW~163--317}\fi
\ifnum#1=5\hbox{OW~158--323}\fi
\ifnum#1=0\hbox{OW~171--334}\fi}
\makeatletter
\title{The spectral index of the ionizing continuum of quasars }
\author{ L. Binette, M. Rodr\'\i guez-Mart\'\i nez and I.
Ballinas\affil{Instituto de Astronom\'{\i}a, UNAM, M\'exico. } }

\fulladdresses{
\item L. Binette, M. Rodr\'\i guez-Mart\'\i nez and I.
Ballinas: Instituto de Astronom\'\i a,
  UNAM,  Ap. 70-264, 04510 M\'exico,
DF, M\'exico (binette@astroscu.unam.mx).}

\shortauthor{Binette et al.}
\shorttitle{The ISED of quasars}

\keywords{atomic processes --- galaxies: intergalactic medium --- 
galaxies: active --- radiative transfer --- ultraviolet: general}

\abstract{%
The ionizing spectral energy distribution of quasars exhibits a
steepening of the distribution short-ward of $1200$~\AA. The change of
the power-law index from approximately $-1$ to $-2$  so far has been
interpreted as being intrinsic to quasars. We study an alternative
interpretation, in which a tenuous absorption screen is responsible
for the change of index. We find that the most successful function of
H{\sc i} with $z$ requires that the ionization fraction be independent
of the metagalactic background radiation, that is frozen in. This
implies that the putative component should be of sufficiently low
density that it could not recombine (while the background is
subsiding) and/or cool from a high temperature state.}

\listofauthors{L. Binette,  M. Rodr\'\i guez-Mart\'\i nez \& I. Ballinas }
\indexauthor{Binette, L.}
\indexauthor{Rodr\'\i guez-Mart\'\i nez, M.}
\indexauthor{Ballinas, I.}

\begin{document}

\maketitle

\section{Introduction} \label{sec:intro}

The ionizing spectral energy distribution (hereafter ISED) of nearby
active galactic nuclei cannot be observed directly, due to the galactic
absorption beyond the Lyman limit. Owing to the redshift effect,
however, we can get a glimpse of the ISED from the spectra of very
distant quasars. 
The pioneering work of Zheng et~al. (1997, ZK97), using HST archived
data, showed that the power-law index ($F_{\nu}\propto \nu^{\alpha}$)
steepens from $\approx -1$ for $\lambda > 1050\,$\AA\ to $\approx -2$
at shorter wavelengths (combining radio-loud and radio-quiet
quasars). Qualitatively similar results were found by Telfer
et~al. 2002 (hereafter TZ02), using a larger sample.

Korista, Ferland \& Baldwin (1997) pointed out that such a steep slope
for the ionizing continuum would imply an insufficient number of
photons beyond $h\nu > 54.4\,$eV to account for the observed
luminosities of the high excitation emission lines.

We report preliminary results of a study aimed at finding an
alternative explanation for the break, one that is {\em extrinsic} to
the quasar spectral energy distribution. We postulate the existence of
a very tenuous absorption gas component that pervades the universe,
and proceed to study the characteristics it must possess to reproduce
the observed break.

\section{Procedure} \label{sec:pro}

The composite spectrum of ZK97 was constructed by merging 284 spectra
of 101 quasars, taken with FOS of the Hubble telescope. The
distribution in quasar redshifts allowed ZK97 to span the wavelength
range of 310--3000\,\AA\ in the quasar rest-frame. Before merging
their spectra, each spectrum was corrected for the Lyman valley and
Ly$\alpha$ forest absorption by calculating the appropriate
transmission curve, using the scheme developed by M\o ller \& Jakobsen
(1990).

Our aim is to reproduce the steepening of the index by introducing a
previously unknown intergalactic absorption component. We hereby
investigate how such a component can reproduce the mean composite
spectrum of ZK97.  Given the exploratory nature of the exercise, we
can either postulate the existence of a very tenuous but continuous
gas component, or a large collection of very thin discrete clouds. At
any rate, these two formalisms would provide an incomplete picture if
the gas distribution resembled that of the Ly$\alpha$ forest
absorption gas (e.g. Dav\'e et~al. 1997).  If the postulated component
consisted of clumps, they must be sufficiently optically thin (and
numerous) so that they do not show up as individually detectable
Ly$\alpha$ absorption lines in the archived FOS spectra, hence
$N_{H^0} \ll 10^{12}\,{\rm cm^{-2}}$.

For definiteness, we adopt a uniform gas distribution. The spectra are
divided in energy bins, and for each quasar rest-frame wavelength bin
$j$, we calculated the transmitted intensity $I_{\lambda_j}^{tr} =
I_{\lambda_j} T_{\lambda_j} = I_{\lambda_j} e^{-\tau(\lambda_j)}$ by
integrating the opacity along the line-of-sight to a quasar of
redshift $z_Q$
\begin{equation}
\begin{array}{cc}
\tau(\lambda_j) = \sum_{i=0}^{10}{\int_{0}^{z_Q}
\sigma_i(\frac{\lambda_j}{1+z})\; n_{H^0}(z) \frac{dl}{dz} \; dz}  \label{eq:tau}
\end{array}
\end{equation}
where $\lambda_j$ is the quasar rest-frame wavelength for bin $j$ and
$n_{H^0}(z)$ the intergalactic neutral hydrogen density. The summation
was carried out over the different opacity sources: photoionization
($i=0$) and line absorption from the Lyman series of hydrogen ($1\le i
\le 10$). Although our code could include up to 40 levels, we found
that considering only the 10 lowest proved to be adequate.  We adopted
a fiducial velocity dispersion $b$ of 30~km/s and assumed a Gaussian
profile for the $\sigma_i$ of the lines.

\section{Calculations}

After various trial and error calculations, experimenting with many
different functional forms for $n_{H^0}$, we found that the following
two distributions of H{\sc i} with redshift could reproduce the
composite spectrum of ZK97 quite well\,:

\begin{displaymath}\label{eq:nha}
\begin{array}{lr}

 n_{H^0}(z) = n^0_{H^0} \, {(z_Q^{\prime})^{0.8}}\, {(1+z)^{1.5}} & (A) 

\end{array}
\end{displaymath}
\begin{displaymath} \label{eq:nhb}
\begin{array}{lr}
 n_{H^0}(z) = n^0_{H^0} \, { ( z_Q^{\prime} \sqrt{z})^{0.7} }\, {(1+z)^{3}}/ \Gamma(z)  & (B) 
\end{array}
\end{displaymath}
where $z$ is the absorbing gas redshift, $z_Q$ the quasar redshift,
$z^{\prime}_Q$ the quasar redshift as seen from the absorbing gas at
$z$ [that is $z^{\prime}_Q = (1+z_Q)/(1+z)\,-1$] and $\Gamma(z)$ is the
parameterized photoionization rate coefficient (Haardt \& Madau
1996) due to the metagalactic ionizing radiation but
renormalized in such a way that $\Gamma(z=0) = 1$\,:

\begin{displaymath}
\begin{array}{c}
\Gamma(z) = (1+z)^{0.73} e^{-0.526 ((z - 2.3)^2 - 2.3^2))} \label{eq:haardt}
\end{array}
\end{displaymath}

The values of $n^0_{H^0}$ at zero redshift are $4.8 \times 10^{-12} \,
{\rm cm^{-3}}$ and $1.35 \times 10^{-11} \, {\rm cm^{-3}}$ for (A) and
(B), respectively.  While the gas is photoionized by the metagalactic
radiation in the case of (B), in (A) the neutral fraction is `frozen
in' since $\Gamma(z)$ is absent. Other factors such as the dependence
on $\sqrt{z}$ in distribution (B), or the exponent of the factor
$(1+z)$ (associated to its evolution during expansion of the
Universe), are not superfluous but essential to the fit (and to the
aim of reproducing the spectral index of $\simeq -2$ beyond
900\AA). Either distribution is quite successful in reproducing the
break as shown in Fig.~1. We have assumed that the intrinsic quasar
distribution exhibits no break and can be described by a single
power-law $F_{\nu} \propto \nu^{-1}$ at all the wavelengths of
interest\footnote{A soft ISED with ${\alpha} = -2$ would appear as an
horizontal line segment in a Log$ F_{\lambda}$ vs. Log$ \lambda$ plot
such as Fig.~1.}. Possible interpretations of either distribution are
not straightforward, due to the presence of factors varying as
$z_q^{\prime}$. It could imply that the gas is anticorrelated with
large baryonic mass accumulation (as represented by the quasars
environment) and is presumably associated with the largest voids.

\begin{figure}
\begin{center}
\includegraphics[width=0.95\columnwidth]{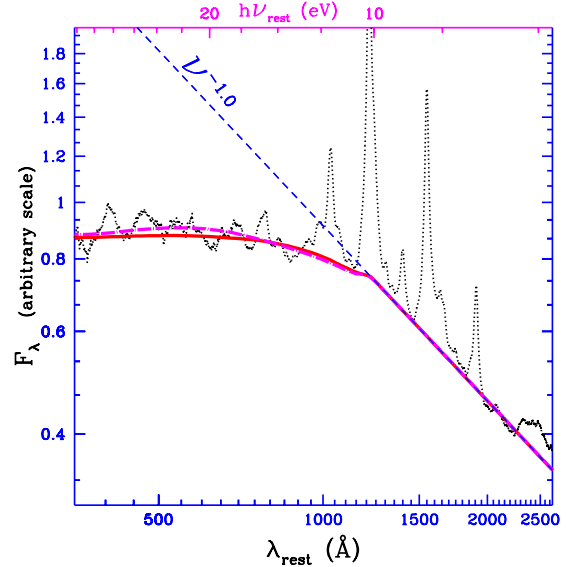}
\end{center}
\caption{The dotted line represents the composite spectrum of  
quasars constructed by Zheng et~al. (1997), adding together both
radio-quiet and radio-loud quasars.  The short-dashed straight line
represents a fit, in the UV range $\lambda > 1200$ \AA, of the
continuum underlying the emission lines ($F_{\nu} \propto
\nu^{-1}$). The solid line is a calculated composite spectrum assuming
a tenuous absorption screen of H{\sc i} as given by distribution
(A). The dotted long-dashed line corresponds to distribution
(B). \label{fig:one} }
\end{figure}

The way in which we simulate the theoretical calculation of a
composite spectrum is by averaging many redshifted spectral distributions
as explained in more detail in Binette et~al. (2002). The
spectrograph wavelength window was assumed to be 3000\,\AA\ to
1300\,\AA\ which we considered typical of the wavelength coverage of
individual quasars by FOS. We redshifted this window in locked steps
to cover the redshift span $0.33 \le z \le 3.6$. Before averaging,
each simulated quasar SED was multiplied by the redshift integrated
transmission curve (c.f. Eqn.~1), taking into account the intervening
gas described by the distribution (A) or (B). We assumed the
concordance $\Lambda$CDM cosmology with $\Omega_{\Lambda} = 0.7$ ,
$\Omega_{M}=0.3$ and $H_0 = 67\,$km/s/Mpc.

\section{Results}

The distribution (B), which takes the metagalactic background
radiation into account, requires the insertion of the extra factor
$\sqrt{z}$ in order to fit the observed ISED. Otherwise, the simulated
spectrum curves down much too steeply at the short wavelength
end. Naively, one would prefer distribution (A) with frozen in
ionization, since it is a simpler expression. A more convincing
argument in favor of (A), however, is the behavior of the power-law
index $\alpha_{EUV}$ of the transmitted ISED in the wavelength range
$\lambda < 900$ \AA. This is shown in Fig.~2 as a function of the
quasar redshift $z_Q$ used in the transmission function. The
distribution (B) shows a significant slope for $\alpha_{EUV}$ while
for (A) the variation is relatively small, confined to the interval
$-1.4 < \alpha_{EUV} < -1.9$.

\begin{figure}
\begin{center}
\includegraphics[width=0.95\columnwidth,viewport={0 0 592 332}]{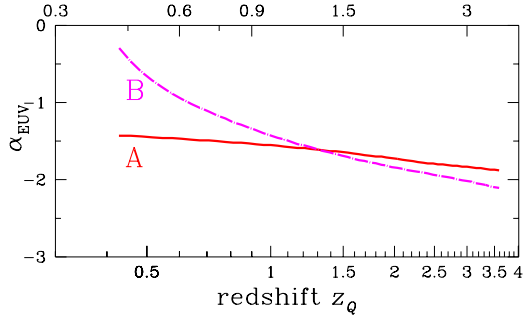}
\end{center}
\caption{Behavior of the  far UV power law-index of the absorbed
ISED, $\alpha_{EUV}$, as a function of quasar
redshift. The solid line and the dotted long-dashed lines correspond to 
assuming distribution (A) and (B), respectively. \label{fig:two}}
\end{figure}

One of the interesting result of TZ02 (c.f. their Fig.~12) is that there is
no global trend of $\alpha_{EUV}$ as a function of redshift. The slope
they found in their linear regression of $\alpha_{EUV}$ vs. Log $z_Q$
was $0.09 \pm 0.56$ and $-0.33 \pm 0.66$ for the populations of
radio-quiet and radio-loud quasars, respectively. Considering the
large dispersion in the $\alpha_{EUV}$ values measured, our slope of
$-0.52$ obtained by linear fit of the solid line in Fig.~2
 cannot be ruled out (we considered radio-loud and
radio-quiet quasars together), while that of  distribution (B) is
undoubtedly inconsistent (giving a linear fit slope of $-1.71$) with the
TZ02 results.

If we substitute $z_Q^{\prime}$ by $z$ in distributions (A) or (B),
the flattening observed in the composite spectrum can be reproduced
although it gives rise to a significant dip at 1216 \AA, which is not
observed. Furthermore, the behavior of $\alpha_{EUV}$ vs. $z_Q$
becomes incompatible with the results of TZ02.

\section{Discussion}

The simpler distribution (A) is in better agreement with the available
data and is marginally consistent with the study of $\alpha_{EUV}$ by
TZ02. Furthermore, it is not plagued by the two problems encountered
by Binette et~al. (2002), who explored different distributions of
$n_{H^0}(z)$ which relied on a long range proximity
effect. Interestingly, a frozen in ionization for the proposed tenuous
absorption component makes sense since it would be similar to that
believed to occur in the Ly$\alpha$ forest at high redshift.

The fact that the frozen in solution is favored is consistent with
the tenuous absorption component being of sufficiently low density
that it did not have time to recombine (as the background subsided)
and/or cool from a high temperature state. If the proposed tenuous
absorption component turned out to be valid, the problem of a too soft
ISED as discussed by Korista, Ferland \& Baldwin (1997) would go away
since the favored intrinsic index of $-1$ used here imply a hard
ionizing spectrum in quasars.

\acknowledgements This work was 
supported by the Mexican science funding agency CONACyT under grant
32139-E. We are indebted to W. Zheng for sharing the original data
presented in ZK97.



\end{document}